\documentstyle[pra,aps]{revtex}
\begin{document}     

\draft

\title
{\bf Galilean Lee Model of the Delta Function Potential}

\bigskip
\author
{C. R. Hagen\cite{Hagen}}

\address
{Department of Physics and Astronomy\\
University of Rochester\\
Rochester, New York 14627-0171}

\maketitle

\begin{abstract}
The scattering cross section associated with a two dimensional delta function
has recently been the object of considerable study.  It is shown here that this
problem can be put into a field theoretical framework by the construction of an
appropriate Galilean covariant theory.  The Lee model with a standard Yukawa
interaction is shown to provide such a realization.  The usual results for 
delta function scattering are then obtained in the case that a stable particle 
exists in the scattering channel provided that a certain limit is taken in the 
relevant parameter space.  In the more general case in which no such limit is 
taken finite corrections to the cross section are obtained which (unlike the 
pure delta function case) depend on the coupling constant of the model. 
\end{abstract}

\bigskip

\begin{center}
{\bf I. Introduction}\\
\end{center}

The problem of scattering by a delta function in two dimensions is of 
considerable interest for a number of reasons, not the least of which is the 
fact that it lacks a dimensional parameter.  This leads directly to the
appearance of divergences in the calculation of bound state energies and
scattering amplitudes, a fact which seriously complicates the task of physical
interpretation.  Although a delta function potential occurs in the relevant 
wave equation for the case of spin one-half Aharonov-Bohm scattering [1], it 
appears there in conjunction with $1/r^2$ terms with coefficients such that a 
cancellation of all divergences occurs.  Since this requires a somewhat 
delicate limiting process (namely, the limit of vanishing flux tube radius 
must be taken at the end of the calculation), it is important to note that no 
such limiting process suffices to yield a finite result for the pure delta 
function potential.  Such a goal can only be achieved by 
a) limiting consideration to the {\it attractive} delta function and 
b) {\it requiring} that there be a bound state associated with the scattering 
channel.  The latter step is frequently justified by pointing out that the 
delta function is so singularly attractive that a bound state is a natural 
expectation.  The crucial point is that a type of renormalization is carried 
out by which divergences are combined into a physical parameter (i.e., the 
bound state energy) in such a way that the relevant scattering amplitude can 
be written as a finite function of the scattering energy and the bound state 
energy.  Just as the physical mass is not amenable to calculation in covariant 
field theory, so also in this case the bound state energy for the delta 
function potential cannot be calculated from first principles.  

This renormalization program has been carried out for the two dimensional delta
function and finite results obtained [2,3].  Bender and Mead [4] have gone one
step further by considering the attractive delta function in $D$-dimensional
space with the $D=2$ result obtained as a limit.  They assert that refs.[2] and
[3] obtain the wrong cross section and infer from this that it is 
{\it essential} that the two dimensional result be obtained as a limit of the 
arbitrary $D$ case.   However, Cavalcanti [5] has recently pointed out that the
results of refs.[2-4] are identical provided only that a calculational error in
ref.[3] is corrected.  The two dimensional delta function thus seems to be 
reasonably well understood within the framework of
conventional Schr\"odinger analysis.

Since Aharonov-Bohm scattering is well known to be the two particle sector of
a Galilean invariant pure Chern-Simons gauge theory, it is natural to ask what
the corresponding Galilean field theory [6] of the delta function should be.  
This paper examines that question and shows that a Yukawa coupling in such a 
theory provides a realization of the delta function potential in the limit in 
which a direct (or contact) interaction is obtained.  In the following section 
the properties of Galilean field theories are briefly reviewed and the Galilean
invariant trilinear interaction term constructed.  The theory obtained from
this process is essentially the Galilean version of the Lee model and has been
discussed previously by L\'evy-Leblond [7].  In section III the two particle
scattering sector is considered and the corresponding Hilbert space
constructed.  This  allows one to calculate the two particle scattering matrix
and thereby obtain a formal expression for the cross section.  In IV the
various limits of the latter are considered and the renormalization carried
out.  The Conclusion summarizes some of the principal results obtained.

\begin{center}
{\bf II A Galilean Model}\\
\end{center}

One begins the construction of an appropriate Galilean covariant model by
the determination of the relevant free particle Lagrangian.  Using the fact
that the invariant quantity in Galilean relativity is $E-{{\bf P^2}\over 2M}$
where $M$ is the particle mass, it is straightforward to infer the free
particle Lagrangian
$${\cal L}_0=\psi^{\dagger} [i{\partial\over \partial t}+ {\nabla^2\over
2M}-U_0]
\psi$$
where $U_0$ is generally referred to as the internal energy parameter.  The
fields $\psi$ and $\psi^\dagger$ satisfy the equal time commutation relation
$$[\psi({\bf x},t),\psi^{\dagger}({\bf x'},t)]=\delta ({\bf x}-{\bf x'})$$
while $\psi$ has the additional property that it annihilates the vacuum state
$|0\rangle$, i.e.,
$$\psi({\bf x},t)|0\rangle=0.$$
It may be noted that no reference is made here to the number of spatial
dimensions although all calculations of scattering amplitudes will consider
only the case of two spatial dimensions.

The construction of the interaction Lagrangian term requires that specification
be made of the particles (or fields) participating in the interaction.  Since a
Yukawa (or trilinear) type of interaction is to be used here, it is convenient
to employ the notation of the Lee model.  The latter considers three particles
$V$, $N$, and $\theta$ with the allowed interactions
\begin{equation}
V\leftrightarrow N+\theta.
\end{equation}
It is to be noted that the Bargmann superselection rule on the mass requires
that the mass parameters of the particles satisfy the relation
$${\cal M}=M+m$$
where ${\cal M}$, $M$, and $m$ denote the masses of the $V$, $N$, and
$\theta$ particles respectively.  One additional remark has to do with the fact
that although the original Lee model considered the $V$ and $N$ fields to be 
fermionic and the $\theta$ to be bosonic, such distinctions are not included 
here since they have no impact on the calculations to be presented.

To determine the most general interaction Lagrangian consistent with Eq.(1) and
the requirement of Galilean invariance one makes use of the general
transformation law
$${\cal U}(g)^{-1}\psi({\bf x},t){\cal U}(g)=exp[iM\gamma(g; {\bf x},t)]
\psi({\bf x'},t')$$
where ${\cal U}(g)$ is the unitary operator associated with the Galilean
transformation
$${\bf x'}=R{\bf x}+{\bf v}t+{\bf a}$$
$$t'=t+b.$$
The parameters $R$, ${\bf a}$, $b$, and ${\bf v}$ refer respectively to
rotations, spatial translations, time translations, and Galilean boosts with
$\gamma(g; {\bf x},t)$ given by
$$\gamma(g; {\bf x},t)={1\over 2}{\bf v}^2t +{\bf v}\cdot R{\bf x}.$$  This
leads to the desired interaction term which is found to be of the form [7]
\begin{equation}
{\cal L}_I({\bf x},t) = 
g_0\int d{\bf y}f(|{\bf y}|)V^{\dagger}({\bf x},t)
 N({\bf x}+{m \over {\cal M}}{\bf y},t)\theta
({\bf x}-{M\over {\cal M}}{\bf y},t)+h.c.
\end{equation} 
where $g_0$ is a coupling constant and $f(|{\bf y}|)$ is a form factor which 
can be arbitrarily prescribed without breaking Galilean invariance.  Although 
the local limit in which $f({\bf y})$ becomes a delta function will eventually
be taken in order to make contact with the delta function potential, it is 
essential to retain for now the regularized version (2).

In order to facilitate the calculation of the scattering matrix it is desirable
to transform all operators to momentum space.  With the definition
$$\psi ({\bf x},t)=\int {d{\bf p}\over 2\pi}e^{i{\bf p\cdot x}}\psi 
({\bf p},t)$$
the nonvanishing commutation relations among the $V$, $N$, and $\theta$ fields 
are of the form
$$[\psi({\bf p},t), \psi^\dagger({\bf p'},t)]=\delta ({\bf p}-{\bf p'}).$$
This leads to the result that the Hamiltonian is 
\begin{equation}
{\cal H}={\cal H}_0+{\cal H}_I
\end{equation}
where [8]
\begin{equation}
{\cal H}_0=\int d{\bf p}[V^\dagger ({\bf p})({{\bf p}^2\over 2{\cal M}}+U_0)V
({\bf p})+N^\dagger({\bf p}){{\bf p}^2\over 2M}N({\bf p})+\theta^\dagger
({\bf p}){{\bf p}^2\over 2m}\theta ({\bf p})]
\end{equation}
and
\begin{equation}
{\cal H}_I=-{g_0\over 2\pi}\int d{\bf P}d{\bf q}[f (\omega)V^\dagger ({\bf P})N
({M\over {\cal M}}{\bf P}+{\bf q})\theta ({m\over {\cal M}}{\bf P}-{\bf q})
+h.c.]
\end{equation}
where $\omega=|{\bf q}|$.  With the derivation of this result the scattering 
amplitude in the $N\theta$ sector can be readily calculated.

\begin{center}
{\bf III The Scattering Matrix}
\end{center}            

As a preliminary to the calculation of the $N\theta$ scattering amplitude       
it should be observed that the Hamiltonian (3-5) implies that the vacuum as well
as the single $N$ and single $\theta$ states are unmodified by the interaction.
Specifically
$${\cal H}|0\rangle=0$$
while
$N^\dagger({\bf p})|0\rangle$ and $\theta^\dagger({\bf p})|0\rangle$ are both
eigenvectors of ${\cal H}$ corresponding to eigenvalues ${\bf p}^2/2M$ and
${\bf p}^2/2m$ respectively.  

The sector which contains an $N\theta$ pair is nontrivial since it is linked by
the interaction to the single $V$ state.  Since one is generally interested in
the case in which the initial configuration of the particles is one of given
momentum, it is appropriate to isolate one term in the expression for the
$N\theta$ state as consisting of $N$ and $\theta$ particles with momenta
${M\over {\cal M}}{\bf P}+{\bf k}$ and ${m\over {\cal M}}{\bf P}-{\bf k}$
respectively.  Thus one writes
\begin{equation}
|{\bf P},{\bf k}^{(+)}\rangle = N^\dagger({M\over {\cal M}}{\bf P}+
{\bf k})\theta^\dagger({m\over {\cal M}}{\bf P}-{\bf k})|0\rangle+\zeta 
V^\dagger({\bf P})|0\rangle+\int d{\bf q}g
({\bf q})N^\dagger ({M\over{\cal M}}{\bf P}+{\bf q})\theta^\dagger 
({m\over {\cal M}}{\bf P}-{\bf q})|0\rangle
\end{equation}
with $\zeta$ and $g({\bf q})$ to be determined from the eigenvalue equation
\begin{equation}
{\cal H}|{\bf P},{\bf k}^{(+)}\rangle = E|{\bf P},{\bf k}^{(+)}\rangle.
\end{equation}
The notation employed here is intended to indicate that the left hand side of
(6) is an outgoing state which in the remote past consisted of an $N\theta$ 
pair at essentially infinite spatial separation.

Upon comparison of the coefficients of the various terms in (7) it is found
that the energy $E$ is given by
$$E={{\bf P}^2\over 2{\cal M}}+{{\bf k}^2\over 2\mu}$$
where $\mu$ is the reduced mass
$$\mu={Mm\over {\cal M}}.$$
It also follows that $\zeta$ and $g({\bf q})$ are related by the equations [9]
$$(U_0-{{\bf k}^2\over 2\mu})\zeta={g_0\over 2\pi}f(\omega_k)+{g_0\over 2\pi}
\int d{\bf q} g({\bf q})f(\omega_q)$$
and
$$g({\bf q})({\bf q}^2-{\bf k}^2){1\over 2\mu}={g_0\over 2\pi}\zeta f^*
(\omega_q).$$
These lead to the result
\begin{equation}
g({\bf q})={({g_0\over 2\pi})^2f^*(\omega_q)f(\omega_k)\over ({{\bf q}^2-{\bf k}
^2\over 2\mu})[U_0-{{\bf k}^2\over 2\mu}-g_0^2\int {d{\bf q}\over (2\pi)^2}
{2\mu |f(\omega_q)|^2\over {\bf q}^2-{\bf k}^2}]}.
\end{equation}
It is to be noted that in writing (8) ${\bf k}^2$ is always to be taken to mean 
${\bf k}^2+i\epsilon$ in accordance with the outgoing wave boundary condition.

The scattering amplitude (8) can be used to construct the scattering matrix
upon noting that incoming states $|{\bf P},{\bf k}^{(-)}\rangle$ are readily
obtained by changing the sign of the $i\epsilon$ term.  Upon adopting a
consistent normalization for two particle states this yields the result for
the scattering matrix $S_{fi}$ in that sector
\begin{eqnarray}
S_{fi}& = & \langle {\bf P}',{\bf k}'^{(-)}|{\bf P},{\bf k}^{(+)}\rangle \\
\nonumber 
& = & (2\pi)^4\delta({\bf P}-{\bf P}')
\left[\delta({\bf k}-{\bf k}')+4\pi i\mu\delta
({\bf k}^2-{\bf k}'^2)
{({g_0\over 2\pi})^2f^*(\omega_{k'})f(\omega_k)\over 
U_0-{{\bf k}^2+i\epsilon\over 2\mu}-g_0^2\int {d{\bf q}\over (2\pi)^2}
{2\mu |f(\omega_q)|^2 \over {\bf q}^2-{\bf k}^2-i\epsilon}}\right].
\end{eqnarray}
The result (9) leads directly to an expression for the cross section in the
form
\begin{equation}
{d\sigma \over d\phi}=(2\pi)^3({{\mu}^2\over k})
\left|{({g_0\over 2\pi})^2f^2(\omega_k)
\over U_0-{{\bf k}^2+i\epsilon\over 2\mu}-g_0^2\int {d{\bf q}\over (2\pi)^2}
{2\mu | f(\omega_q)|^2\over {\bf q}^2-{\bf k}^2-i\epsilon}}\right| ^2.
\end{equation}
Although one clearly is most interested in the local limit of (10) (i.e., the
case in which $f(\omega)$ is constant), the resulting divergence in the
integral over ${\bf q}$ in that limit means that that limiting procedure must be
handled with some caution.  This is the renormalization process to which
attention is now directed.

\begin{center}
{\bf IV Renormalization and the Delta Function Limit}
\end{center}

In order to deal with the issue of how to take the local limit $f(\omega)=1$ it
should be noted that the relevant quantity is 
$$G_V^{-1}(U)=U_0-U -g_0^2\int {d{\bf q}\over (2\pi)^2} {2\mu |f(\omega_q)|^2
\over {\bf }q^2-2\mu U}$$
where 
$$U\equiv E-{{\bf P}^2\over 2 \cal M}$$
which also happens to be the inverse propagator for the $V$ particle.  The 
application of the usual renormalization procedure requires either that the 
inverse propagator vanish on the physical sheet in the case of a stable (i.e., 
bound state) particle or that its real part vanish for the case of a resonance. 
Since this can happen in the local limit only in the case $U_0>0$, it is 
henceforth assumed that this condition is satisfied [10].  One also observes 
from
the form of $G_V^{-1}(U)$ that the behavior of the integrand which appears in
that quantity guarantees that the bound state condition will be satisfied for
some $U<0$.  Taking this value of $U$ to be $-E_0$ there follows that 
$$U_0=-E_0+g_0^2\int {d{\bf q}\over (2\pi)^2}{2\mu |f(\omega_q)|^2\over {\bf q}
^2+2\mu E_0}.$$
Using this result to eliminate $U_0$ from $G_V^{-1}(U)$ one obtains the
renormalized form 
$$G_V^{-1}=-(U+E_0)\left[1+g_0^2\int {d
{\bf q}\over (2\pi)^2}{2\mu \over 
({\bf q}^2-2\mu U)({\bf q}^2+2\mu E_0)}\right]$$
where the local limit has been taken in the convergent integral over $q$.

This result upon insertion in (10) yields the differential cross section as 
$${d\sigma\over d\phi}={2\pi\over k}\left|-i\pi+log{k^2\over 2\mu E_0}-
{{k^2\over 2\mu} + E_0\over \mu g_0^2/2\pi}\right|^{-2}$$
and the total cross section result
\begin{equation}\sigma={4\pi^2\over k[\pi^2+(log{k^2\over 2\mu E_0}-
{{k^2\over 2\mu}+E_0\over \mu g_0^2/2\pi})^2]}.
\end{equation}
This can also be expressed in term of a renormalized coupling constant $g$
with $(g/2\pi)^2$ defined as the residue of the scattering amplitude at the 
pole.  Thus 
$$g^2={g_0^2\over 1+{\mu g_0^2\over 2\pi E_0}}$$
or, equivalently,
$$g_0^2={g^2\over 1-{\mu g^2\over 2\pi E_0}}.$$

The result (11) coincides in the limit $g_0\to \infty$ with the results
obtained previously for the delta function [2-5].  This is, of course, expected
since the equation of motion for $V({\bf x},t)$ implied by the Hamiltonian 
(3-5) is of the form
$$(i{\partial\over \partial t} + {\nabla^2\over 2\cal M}- U_0)V({\bf x},t)=
- g_0\int d{\bf y}f(|{\bf y}|)N({\bf x}+
{m\over \cal M}{\bf y},t)\theta({\bf x}-
{M\over \cal M}{\bf y},t)$$
and has the property that for $U_0,g_0\to\infty$ with ${g_0^2 \over U_0}$
fixed, it becomes an equation of constraint.  Thus Eq.(2) is in that limit
\begin{equation}
{\cal L}_I=\lambda \left|\int d{\bf y}f(|{\bf y}|)
N({\bf x} + {m\over \cal M} {\bf y},t)\theta({\bf x}-
{m\over \cal M}{\bf y},t)\right|^2
\end{equation}
where $\lambda>0$ is defined by
$$\lambda\equiv {g_0^2\over U_0}.$$
Using the approach of III it follows that Eq.(12) yields the usual momentum
space equations for the delta function potential, thereby establishing that the
above limit is indeed the field theory of the delta function potential.  

\begin{center}
{\bf V Conclusion}
\end{center}

In this work a Galilean field theory of the two-dimensional delta function has
been presented.  A noteworthy extension of the delta function problem has also
been achieved by using as the framework for this discussion a trilinear or 
Yukawa coupling.  The explicit and exact expression for the S matrix has been 
derived in the two particle sector of the model and subsequently used to derive
the cross section.  It was found that while the cross section for this extended
model has correction terms to the delta function potential, there is agreement 
in the limit in which a contact (or quadrilinear) coupling is obtained.  The
renormalization process has been seen to play an essential role in achieving 
the goal of taking the local (i.e., delta function) limit.  

\acknowledgments

This work is supported in part by the U.S. Department of Energy Grant
No.DE-FG02-91ER40685.



\begin{references}
\bibitem{  } Of course one could also construct an appropriate theory in 
Minkowski space.  Since, however, this would require at some point that a 
nonrelativistic limit be taken, it is clearly desirable to build in that limit 
at the outset (i.e., to construct a Galilean theory).
\bibitem{ } It should be noted that the bare (i.e., unrenormalized) internal
energies of the $N$ and $\theta$ fields have been taken to be zero.  That this
can be done without loss of generality is a consequence of the fact that $\psi$
can always be replaced by $\psi e^{-iU_0t}$,  thereby eliminating the internal
energy term for each of these fields.  That this cannot also be done for the 
$V$ field is a consequence of the fact that the interaction term constrains the
set of allowable gauge transformations.
\bibitem{ } It should be noted that there is a sign error in the corresponding
calculation of ref.[7].
\bibitem{ }  Alternatively stated, the local field theory with $U_0<0$ is a
null theory.
\end{references}
\end{document}